\documentclass[12pt]{article}
\begin{document}
\vbadness = 100000
\hbadness = 100000

\title{Superluminal neutrinos}
\author{Jerrold Franklin\footnote{Internet address:
Jerry.F@TEMPLE.EDU}\\
Department of Physics\\
Temple University, Philadelphia, PA 19122-6082
\date{\today}}
\maketitle
\begin{abstract}
The superluminal propagation of neutrinos observed by the OPERA collaboration can be explained by an energy dependent potential for the neutrino beam in passage through the Earth. 

\end{abstract}

\section{Introduction}

Since the publication of evidence of superluminal propagation of muon neutrinos by the OPERA collaboration\cite{opera}, there have been a number of theoretical papers[2-52] providing various scenarios for superluminal propagation.  In this note, we introduce a new proposal that gives a natural explanation that is consistent with special relativity and is within the usual standard model.  Most of the 730 kilometer path of the neutrinos observed by OPERA is underground.  We show below that an energy dependent potential acting on the neutrinos in their passage through matter can lead to superluminal propagation.

The OPERA collaboration finds an early arrival time  
 for neutrinos faster than if they traveled at the speed of light $c$.  This corresponds to a relative neutrino speed given by
\begin{equation}
\Delta=v/c-1=c\delta t/L,
\end{equation}
where $L$ is the 730 km propagation distance, and  $\delta t$ is the difference between the time of flight for light minus the neutrino time of flight.
The arrival times were found for two different neutrino energy ranges, $E_1=13.9$ GeV and $E_2=42.9$ GeV.
The measured time differences were 
\begin{eqnarray}  
\delta t_1& = & 53.1\pm 18.8\pm7.4\,{\rm ns}
\label{eq:t1}\\
\delta t_2& = & 67.1\pm 18.2\pm7.4\,{\rm ns},
\label{eq:t2}
\end{eqnarray}
where the first error is statistical and the second systematic.
These early arrival times
lead to superluminal relative neutrino velocities of
\begin{eqnarray}  
\Delta_1& = & (2.18\pm .77\pm .30)\times 10^{-5}\quad (E_1=13.9\,{\rm GeV})
\label{eq:d1}\\
\Delta_2& = & (2.76\pm .75\pm .30)\times 10^{-5}\quad (E_2=42.9\,{\rm GeV}).
\label{eq:d2}
\end{eqnarray}
Within the relatively large statistical errors, these relative velocities are consistent with no energy dependence, although there could be some variation with energy.

Superluminal speeds are not forbidden by special relativity.  What is forbidden is the acceleration of a particle of fixed invariant mass $m$ from subluminal to superluminal speed.  This is because the equation
\begin{equation}
E=\frac{mc^2}{\sqrt{1-v^2/c^2}}
\label{eq:emc}
\end{equation}
shows that infinite energy would be required for a particle of fixed mass to approach the speed of light.  

However, particles that are produced at superluminal speeds are consistent with relativity, 
and numerous theoretical papers have been written considering that possibility.
A general class of particles with negative mass squared, $m^2<0$, called tachyons, would have energy dependent superluminal speeds with\cite{drago} 
\begin{equation}
\Delta=\frac{-m^2c^4}{(1+v/c)E^2}\simeq -m^2 c^4/2E^2.
\label{eq:tach}
\end{equation} 

As Ref.\cite{drago} points out, this class of superluminal propagation is ruled out by the lack of strong energy dependence in the OPERA arrival times, since 
Eq.\ (\ref{eq:tach}) would predict that the ratio $\Delta_1/\Delta_2$ would be 9.
In fact any particle of fixed mass produced at superluminal speed in vacuum would have the energy dependence of $\Delta$ as in eq. (\ref{eq:tach}) because of the Lorentz invariance of $E^2-p^2 c^2$.  This leads us to consider that the neutrinos are produced with subluminal speed, but propagation through matter produces superluminal propagation.  We demonstrate this in the following section.

\section{Superluminal propagation in matter}

We consider a neutrino of mass $m$ whose motion is described by a Dirac equation that includes a potential $V$ that is independent of position, but does have energy dependence:
\begin{equation}
[{\bf\alpha\cdot p}+\beta m +V]\psi=E\psi,
\end{equation} 
where {\boldmath$\alpha$\unboldmath} and $\beta$ are the usual Dirac matrices, and we are using units with $c=1$ and $\hbar=1$.
The four component wave function $\psi$ can be written in terms of two component spinors $u$ and $v$ as
\begin{equation}
\psi =  N\left(\begin{array}{c}u\\ v\end{array}\right),
\end{equation}  
with $N$ an appropriate normalization constant.
The two component spinors $u$ and $v$ satisfy the equations
\begin{eqnarray}  
({\bf\sigma\cdot p})u & = & (E-V+m)v
\label{eq:u}\\
({\bf\sigma\cdot p})v & = & (E-V-m)u. 
\label{eq:v}
\end{eqnarray}

Solving Eq. (\ref{eq:u}) for $v$ and substituting into Eq. (\ref{eq:v}) gives
\begin{equation}
p^2 u=[(E-V)^2-m^2]u=[E^2-2EV+V^2-m^2]u.
\label{eq:p2}
\end{equation}
The group velocity of the neutrino wave is given by
\begin{equation}
v_G=\frac{dE}{dp}=\frac{p}{E(1-V/E-V'+VV'/E)},
\label{eq:vg}
\end{equation}
where $V'=dV/dE$.

We assume that $m<<V<<E$, and find $p$ to lowest order in $V/E$:
\begin{equation}
p=\sqrt{E^2-2EV+V^2-m^2}\simeq E(1-V/E).
\end{equation} 
Then, Eq.\ (\ref{eq:vg}) can be written as
\begin{equation}
v_G=\frac{1-V/E}{1-V/E-V'}\simeq 1+V'.
\label{eq:vg2}
\end{equation}
If the potential $V$ is proportional to the energy,
\begin{equation}
V=\mu E,\quad V'=\mu,
\label{eq:ve}
\end{equation}
and the OPERA parameter $\Delta$ is given by
\begin{equation}
\Delta=v_G-1=V'=\mu.
\label{eq:dvg}
\end{equation}

We see that the neutrino velocity is superluminal, and the degree of superluminosity is independent of the energy, as indicated by OPERA.
More OPERA data to lower the statistical errors could give a better measure of the energy dependence.  A slow energy dependence could be accommodated by a fractional  change in the exponent of the energy dependence of $V$.  

\section{SN1987a}

The mechanism we propose for superluminal  neutrino propagation can resolve two puzzles posed by the neutrinos observed from the supernova SN1987a.
A total of 26 neutrinos were observed in the Kamiokande-II, IMB, and Baksan detectors, with a spread in arrival times of about 15 seconds. 
All arrived close to the first appearance of the light signal.  Because of the large distance ($1.7\times 10^5$ ly) to the supernova, the superluminal speed indicated by 
Eqs.\ (\ref{eq:t1}) and (\ref{eq:t2}) would have the neutrinos arriving years before the light signal, with a large spread in the individual arrival times.  This seeming incompatibility between the OPERA results and the SN1987a neutrino arrival times does not arise if the superluminal propagation is due to neutrino interaction with matter, because there is very little matter in outer space.

The small amount of matter that does occur in space could resolve the other, long standing, puzzle about the 1987a neutrinos.
In addition to the 26 neutrinos observed by the Kamiokande-II, IMB, and Baksan detectors, 5 neutrinos were observed at the  LSD detector under Mont Blanc.
BUT, these neutrinos arrived 5 hours before the other neutrinos.  Because of this inexplicable early arrival, the tendency has been to discount these neutrinos as unrelated to SN1987a.   The only occurrence ever of 5 neutrinos in the LSD detector within several seconds of each other, so close to the main neutrino shower, has been considered a (unlikely) coincidence.  

The early appearance of the LSD neutrinos is a good candidate for superluminal propagation arising from the interaction of 
the neutrino beam with the small amount of matter in its path.
The fast LSD neutrinos would correspond to the lowest neutrino mass eigenstate.  If this were the only neutrino mass lighter than the potential $V$ in outer space, it would explain why the LSD neutrinos arrived ahead of the other neutrinos.  The 5 hour lead time  in the $1.7\times 10^{5}$ year journey from SN1987a corresponds to $\Delta=3.3\times 10^{-9}$, which is much smaller than the $2.5\times 10^{-5}$ for the OPERA neutrinos. This is to be expected because of the lesser density of matter. This makes the effect of the potential $V$ a reasonable explanation for the early appearance of the  LSD neutrinos.

\section{Conclusion}

We have made several assumptions and simplifying approximations, but have demonstrated that interaction of a neutrino beam in matter via an energy dependent potential 
can produce superluminal propagation as observed by the OPERA collaboration.  This is expected only for neutrinos, whose mass can be smaller than the potential causing the superluminosity.

\end{document}